\documentclass[runningheads,a4paper]{llncs}

\usepackage{eso-pic}
\AddToShipoutPicture*{
\put(133,160){
\scriptsize
Proceedings of WLPE 2013, {\tt arXiv:1308.2055}, August 2013.
}}

\usepackage{dsfont}
\usepackage{amsmath}
\usepackage{amssymb}
\usepackage{amsfonts}
\usepackage{latexsym}
\usepackage{float}
\usepackage[pdftex]{hyperref}
\usepackage{algorithm}
\usepackage{algorithmic}
\usepackage{verbatim}
\usepackage{listings}
\hypersetup{colorlinks=true}
\usepackage{color}
\usepackage{microtype}
\usepackage{graphicx}
\usepackage{ngrammar}
\usepackage{multirow}
\usepackage{makeidx}  
\usepackage{url}
\urldef{\mails}\path|{shadjichrist,porter,warren}@cs.stonybrook.edu|
\newcommand{\keywords}[1]{\par\addvspace\baselineskip
\noindent\keywordname\enspace\ignorespaces#1}

\begin{document}
\mainmatter
\title{Efficiently Retrieving Function Dependencies in the Linux
  Kernel Using XSB}


\author{Spyros Hadjichristodoulou \and Donald E. Porter \and David S. Warren}

\institute{Computer Science Department \\ Stony Brook University \\
  Stony Brook, NY 11794-4400 \\
\mails\\}


\ifx \oscardebug \undefined
\newcommand{\fixmedp}[1]{{}}
\else
\newcommand{\fixmedp}[1]{{\bf\textcolor{red}{ [ dP FIXME: #1 ]}}}
\fi

\maketitle

\def\longconferencenames{}


\newcommand{\conferencename}[3]{
\ifx\longconferencenames\undefined
\newcommand{#1}[0]{{#2}}
\else
\newcommand{#1}[0]{{#3}}
\fi
}

\newcommand{\cvnote}[1]{
\ifx\includecvnotes\undefined
\else
{#1}%
\fi
}

\newcommand{\bionote}[1]{
\ifx\includebionotes\undefined
\else
{#1}%
\fi
}

\conferencename{\podc}{PODC}{Proceedings of the ACM symposium on Principles of
distributed computing (PODC)}

\conferencename{\asplos}{ASPLOS}{{Proceedings of the ACM International Conference on Architectural Support for Programming Languages and Operating Systems (ASPLOS)}}

\conferencename{\spaa}{SPAA}{{Proceedings of the ACM symposium on Parallelism in algorithms and architectures (SPAA)}}

\conferencename{\osdi}{OSDI}{{Proceedings of the USENIX Symposium on Operating Systems Design and Implementation (OSDI)}}

\conferencename{\disc}{DISC}{{Proceedings of the International Conference on Distributed Computing (DISC)}}

\conferencename{\usenixatc}{USENIX}{{Proceedings of the USENIX Annual Technical Conference}}
\conferencename{\usenixsec}{USENIX Security}{{Proceedings of the USENIX Security Symposium}}

\conferencename{\pldi}{PLDI}{{Proceedings of the ACM SIGPLAN conference on Programming language design and implementation (PLDI)}}

\conferencename{\computer}{Computer}{{IEEE Computer}}

\conferencename{\sosp}{SOSP}{{Proceedings of the ACM SIGOPS Symposium on Operating Systems Principles (SOSP)}}

\conferencename{\isca}{ISCA}{{Proceedings of the ACM IEEE International Symposium on Computer Architecture (ISCA)}}

\conferencename{\csaw}{CSAW}{{Proceedings of the ACM Workshop on Computer Security Architecture (CSAW)}}

\conferencename{\wddd}{WDDD}{{Proceedings of the Workshop on Duplicating, Deconstructing, and Debunking (WDDD)}}

\conferencename{\vldb}{VLDB}{{Proceedings of the International Conference on Very Large Databases (VLDB)}}

\conferencename{\toplas}{TOPLAS}{{ACM Transactions on Programming Languages and Systems (TOPLAS)}}

\conferencename{\tocs}{TOCS}{{ACM Transactions on Computer Systems (TOCS)}}

\conferencename{\ppopp}{{PPoPP}}{{Proceedings of the ACM SIGPLAN Symposium on Principles and Practice of Parallel Programming (PPoPP)}}

\conferencename{\jpdc}{J. Parallel Distrib. Comput.}{{Journal of Parallel and Distributed Computing}}

\conferencename{\ismm}{ISMM}{{Proceedings of the ACM International Symposium on Memory Management (ISMM)}}

\conferencename{\cacm}{CACM}{{Communications of the ACM (CACM)}}

\conferencename{\hpca}{HPCA}{{Proceedings of the IEEE International Symposium on High-Performance Computer Architecture (HPCA)}}

\conferencename{\transact}{TRANSACT}{{Proceedings of the ACM SIGPLAN Workshop on Transactional Computing (TRANSACT)}}

\conferencename{\iiswc}{IISWC}{{Proceedings of the IEEE International Symposium on Workload Characterization (IISWC)}}

\conferencename{\tpds}{IEEE Trans, Parallel Distrib. Syst.}{{IEEE Transactions on Parallel and Distributed Systems}}

\conferencename{\osr}{OSR}{{ACM Operating Systems Review}}

\conferencename{\nsdi}{NSDI}{{Proceedings of the USENIX Symposium on Networked Systems Design and Implementation (NSDI)}}

\conferencename{\cc}{CC}{{Proceedings of the International Conference on Compiler Construction (CC)}}

\conferencename{\surveys}{ACM Comput. Surv.}{{ACM Computing Surveys}}
\conferencename{\icde}{ICDE}{{Proceedings of the IEEE International Conference on Data Engineering (ICDE)}}
\conferencename{\fast}{FAST}{{Proceedings of the USENIX Conference on File and Storage Technologies (FAST)}}
\conferencename{\eurosys}{{E}uro{S}ys}{{Proceedings of the ACM European Conference on Computer Systems ({E}uro{S}ys)}}
\conferencename{\hotos}{HotOS}{{Proceedings of the USENIX Workshop on Hot Topics in Operating Systems (HotOS)}}
\conferencename{\hotcloud}{HotCloud}{{Proceedings of the USENIX Workshop on Hot Topics in Cloud Computing (HotCloud)}}
\conferencename{\oopsla}{OOPSLA}{{Proceedings of the ACM SIGPLAN Conference on Object-Oriented Programming, Systems, Languages, and Applications (OOPSLA)}}
\conferencename{\ndss}{NDSS}{{Proceedings of the Network and Distributed System Security Symposium (NDSS)}}
\conferencename{\oakland}{Oakland}{{Proceedings of the IEEE Symposium on Security and Privacy (Oakland)}}
\conferencename{\ispass}{ISPASS}{Proceedings of the IEEE International Symposium on Performance Analysis of Systems and Software (ISPASS)}

\conferencename{\europar}{{E}uro{P}ar}{{Proceedings of the European Conference on Parallel Programming ({E}uro{P}ar)}}

\conferencename{\sigcse}{{SIGCSE}}{{Proceedings of the ACM SIGCSE technical symposium on Computer science education (SIGCSE)}}

\conferencename{\ccs}{{CCS}}{{Proceedings of the ACM Conference on Computer and Communications Security (CCS)}}

\conferencename{\veeconf}{{VEE}}{{Proceedings of the International Conference on Virtual Execution Environments (VEE)}}

\conferencename{\lisa}{{LISA}}{{Proceedings of the Large Installation System Administration Conference (LISA)}}
\conferencename{\scool}{SCOOL}{{Proceedings of the Workshop on Synchronization and Concurrency in Object-Oriented Languages (SCOOL)}}
\conferencename{\cgo}{CGO}{{Proceedings of the International Symposium on Code Generation and Optimization (CGO)}}
\conferencename{\dsn}{{DSN}}{Proceedings of the International Conference on Dependable Systems and Networks (DSN)}
\conferencename{\sac}{{SAC}}{{Proceedings of the ACM Symposium on Applied Computing (SAC)}}

\conferencename{\cluster}{{IEEE Cluster}}{{IEEE International Conference on Cluster Computing}}

\begin{abstract}
  In this paper we investigate XSB-Prolog as
  a static analysis engine for data represented by medium-sized
  graphs. 
  We use XSB-Prolog to automatically identify 
  function dependencies in the Linux Kernel---queries that are difficult
  to implement efficiently in a commodity database and that developers
  often have to identify manually. 
  This project illustrates that Prolog systems are ideal for 
  building tools for use in other disciplines
  that require sophisticated inferences,
  because Prolog is both declarative and can efficiently implement
  complex problem specifications through tabling and indexing.

\end{abstract}

\section{Introduction}
When the idea of Logic Programming was conceived in the early 
1970s~\cite{colmerauer1996birth}, 
its primary application
was Natural
Language Processing (NLP)~\cite{covington1994natural}. The purely
declarative nature of Prolog programs allows programmers to 
specify a problem's requirements,   
and then leave the Prolog engine to
actually solve it. Over the past years, researchers 
have developed a range of Prolog optimizations, including
indexing~\cite{rao1997xsb} and incremental
tabling~\cite{saha2006,saha2005symbolic}, which
make Prolog-generated solutions efficient.
Because programmers can easily specify solutions in Prolog, and because
the generated solutions are efficient, 
it has the potential to be a practical tool for general-purpose
problem solving.
However, Prolog has not been widely adopted outside Artificial
Intelligence (AI); it is traditionally employed for applications in expert system
building~\cite{merritt1989building}, ontology
representation~\cite{laera2004sweetprolog,papadakis2011proton} 
and theorem proving~\cite{stickel1988prolog}. One of the most 
notable and recent example of an application of Prolog in NLP is the
implementation of IBM's Watson computer system~\cite{watson}. In 2011, Watson
competed against former American television ``Jeopardy'' game winners
in answering questions posed in natural language and won the grand
prize\footnote{\url{http://www.computerworld.com/s/article/9209938/Watson_triumphs_in_em_Jeopardy_em_s_man_vs._machine_challenge}}. 

Logic Programming has been adopted by industry as well,
primarily to make 
complex inferences over a data 
set~\cite{huang2011datalog,dlv,semmle,ramakrishnan2007xcellog}, 
or as a security policy specification language~\cite{becker2006secpal}. 
Despite these advances showing the utility of Logic Programming
for developing robust software systems, many
opportunities for further adoption remain. We believe that Prolog's
combination of easy-to-specify 
solutions 
with the efficient implementations
makes Prolog ideal for use in other CS disciplines
as well, especially where analysis, knowledge representation, and
inference over large amounts of data is needed. 

One example of a CS
discipline where such analyses are needed is  Operating Systems (OS). Operating
systems are usually written to support a wide range of hardware, including different
instruction set architectures, 
and a single OS is often several million lines of code written in
multiple languages.  
For instance, the Linux Kernel has components written in C and
various assembly dialects, and its total
code base exceeds 15,000,000 lines of code. What makes understanding
the Linux Kernel even more difficult is its complex
set of compilation options, which are in turn implemented by heavy use
of C preprocessor macros and, in some cases, multiple versions of the
same function---all of which can obfuscate the code and frustrate
simple text searches.

There are many development tasks that require an expert to understand
and manually reason about 
this large body of code.
For instance, certain synchronization primitives in Linux, such as spinlocks
and read-copy update (RCU), require that a blocking function not 
be called while in a critical section.  
In order to add a function call inside a critical section
without violating this invariant,
one must essentially determine whether {\tt schedule()} 
is in the transitive closure of all functions
that could be called by a newly-added line of code.

The Linux Cross-Reference
(LXR\footnote{\url{http://lxr.linux.no/}}) is a tool that helps
developers understand the Linux kernel source. 
Because the LXR is
implemented using a traditional RDBMS, PostgreSQL in particular, the RDBMS
can efficiently execute simple queries, such as locating 
all instances of a particular string in the code.
However, a traditional RDBMS
does not offer any kind of \textit{reasoning} or \textit{inference} over the
data.
In our example of finding whether {\tt schedule()} is in the transitive closure
of a function call, answering this query in a traditional RDBMS
would require loading all of the tables in memory, and then joining the tables
using the callee function's name as the key. This would be too
inefficient for databases containing information about huge code bases
such as the Linux Kernel. Further, handling the case
where function \texttt{A} calls a chain of intermediate functions that 
ultimately call function \texttt{B} further increases these overheads
in a traditional RDBMS.

Queries that require such inferences or deductions can be 
implemented using a \textit{deductive database}---a database
optimized for deductions over large data sets.
XSB-Prolog  is among the most efficient deductive database systems
available~\cite{sagonas1994xsb}, hence it is ideal for inferring
information from medium and large datasets.

This paper describes the design and preliminary experiences 
using XSB-Prolog to build a tool to help developers reason about the
Linux kernel source code. 
This paper focuses on the transitive closure problem described above;
we are extending the tool as ongoing work.
Our tool is available at
\url{http://ewl.cewit.stonybrook.edu/spyros/kernel.php}. 
Section~\ref{eff} presents 
space and time measurements, demonstrating that
tabling and indexing in Prolog systems make the difference 
between practical and impractical tools.

\section{Description of the Problem}
As any large code base evolves, developers may discover the need to
modularize and reuse 
functionality.  For instance,
file systems often "reinvent the wheel" in developing similar API features
or techniques
for managing consistency across metadata writes.
Linux supports dozens of file systems that offer various features and performance characteristics;
although some components are shared (e.g., the Linux {\tt libfs} and the 
{\tt ext3} {\tt jbd} journal), separating a feature into a reusable module 
is a manual process only undertaken by an expert~\cite{ext3}.
As a result, once a feature has proved useful in one file system, the feature is not 
easily adapted to all other file systems.
Thus, useful features languish in individual file systems or research prototypes, 
such as transactions~\cite{TxF,spillane09fast},
atomic append~\cite{googlefs} and copy-on-write checkpointing~\cite{btrfs}.

A key question a developer must answer when modularizing code is essentially:
where is the most functionally narrow point in this code base at which to create
a shared API?  Or, for any given line of code or function call used in the implementation of a feature, 
does it make more sense to bring along supporting code?
As another example of this issue, consider porting a data structure from 
a user-level library into the OS kernel: for each library call the data structure makes,
should the developer copy in that library function, adapt to a similar function provided in the kernel,
or re-engineer that part of the code to avoid the use of the library call?
These design decisions can be subtle, and the designer 
could benefit greatly from a tool that automatically identifies
how difficult a given function is to excise from its supporting code base.

Finally, even more mundane tasks require OS kernel developers to 
reason about the transitive closure of all possible functions a given function could call.
For instance, if a developer is modifying a function that acquires a lock, the developer must not
call any functions that could acquire a second lock that violates the kernel's global lock ordering---requiring
 the developer to understand all possible code paths or risk introducing deadlocks.
Similarly, read-copy update (RCU)~\cite{mckenney04rcu} is designed with the invariant that a reader will not 
call a blocking function inside a critical section; this again requires a deep understanding
about all possible disk reads, network accesses, memory allocations, etc.
Although Linux can compile in dynamic checks that can catch these bugs, these tools will only work 
if the checks are correctly written and all code paths are tested.
Developers could avoid ever introducing these subtle bugs if they had 
the ability to double-check these global invariants while writing the code.

\section{Our Solution}\label{solution}
A key observation of this work is that the power of a tool to help users 
make inferences about a large dataset is 
determined by the power of the underlying DBMS.
Because LXR is built using a traditional RDBMS,
it cannot support even simple queries that require 
recursion.
Trying to approximate recursive queries in PostgreSQL
would
require multiple joins of large tables. 
In XSB-Prolog
however, solutions to such problems are both easy to specify and efficient, 
because tabling ensures termination and efficiency 
of the various transitive closure definitions
(left-recursive, right-recursive, and doubly-recursive).
Our aim is to facilitate 
issuing 
simple queries with sophisticated implementations
by OS developers and researchers.
We further aim to encapsulate the details of how Prolog works
and how these queries are implemented from the users.
Even if a
sophisticated engine is required to answer these queries, the
user interface should be simple and intuitive.

Thus, the following key
components are required for the development of our tool, as explained
in the following sections:
\begin{enumerate}
\item Extracting function dependency information from the Linux
  Kernel.
\item Representing the function dependencies in a way that is 
  easily processed by Prolog systems.
\item Designing an easy-to-use interface for this tool, accessible even
  for users who are not proficient in Prolog. 
\end{enumerate}

\section{Extracting Function Dependency Information from the Linux
  Kernel} 

\begin{figure}[t!b!]
\begin{lstlisting}[language=C]
#ifdef CONFIG_STACK_GROWSUP
int expand_stack(struct vm_area_struct *vma, unsigned long address)
{
        struct vm_area_struct *next;

        address &= PAGE_MASK;
        next = vma->vm_next;
        if (next && next->vm_start == address + PAGE_SIZE) {
                if (!(next->vm_flags & VM_GROWSUP))
                        return -ENOMEM;
        }
        return expand_upwards(vma, address);
}
#else 
int expand_stack(struct vm_area_struct *vma, unsigned long address)
{
        struct vm_area_struct *prev;

        address &= PAGE_MASK;
        prev = vma->vm_prev;
        if (prev && prev->vm_end == address) {
                if (!(prev->vm_flags & VM_GROWSDOWN))
                        return -ENOMEM;
        }
        return expand_downwards(vma, address);
}
#endif
\end{lstlisting}
\caption{An excerpt from {\tt mm/mmap.c} in Linux 3.10, illustrating
  the use of the C preprocessor to select between two different implementations 
  of a function.}
\label{fig:expand}
\end{figure}

In order to extract function dependency information from the
Linux Kernel, we used \texttt{GCC} and
\texttt{egypt}\footnote{\url{http://www.gson.org/egypt/egypt.html}}.
The \texttt{egypt} tool
is a Perl script that parses the intermediate code
representation of C source files and outputs a
relevant \texttt{Graphviz}\footnote{\url{http://www.graphviz.org/}}
file, which can be used to graphically represent the call
graph. 

The \texttt{egypt} tool takes the compiler's intermediate code representation
as input; to output this intermediate representation,
we compile the source code using the 
\texttt{GCC} using the \texttt{fdump-rtl-expand} compilation flag.
The only change to the Linux kernel is adding this flag to the makefile.
By compiling the entire kernel with this extra flag, we get one extra output
file per C source file with extension \texttt{.c.144r.expand}. These
files contain intermediate code information in the form of \textit{Register
Transfer Language} (RTL).

Running \texttt{egypt} on
each of these RTL files outputs call-graph information in an
easy-to-read manner, i.e. in the form of \texttt{Function A ->
  Function B} expressions. The Linux Kernel version we used is
\textbf{3.6.6}, and we performed an \texttt{allnoconfig}
compilation, which is a minimal configuration that disables all
optional features. 

One drawback of our current design is that, 
because we extract dependencies after the preprocessor runs,
we cannot easily capture function dependencies that might arise in a 
different configuration. Recall that the Linux Kernel provides many compilation options,
which are generally selected using C preprocessor macros.
For example, whether a program stack adds frames at a 
higher or lower virtual address (grows "up" or "down") 
is controlled by a compile-time option.  
Figure~\ref{fig:expand} lists an excerpt of {\tt mm/mmap.c} from Linux 3.10, which
illustrates the 
potential to miss possible dependencies.

Because our current prototype identifies function dependencies
after {\tt CONFIG\_\-STACK\_\-GROWSUP} is evaluated,
it can either miss {\tt expand\_\-upwards} or {\tt expand\_\-down\-wards}
as a potential dependency.
We are currently investigating ways to retain the simplicity of RTL 
without losing information about the preprocessor configuration directives.

\section{Easy-to-process Representation of Information}
This section describes how we process the output of \texttt{egypt}
using XSB-Prolog, and then use this information to assert facts
describing the call graph into the database.
The output \texttt{.eg} files generated 
use a fairly small subset of the \texttt{Graphviz} language,
specified by the grammar below:

\begin{ngrammar}[notation=ebnf]
  \nsdef{graph}{%
    \nst{digraph}\nst{callgraph}\nst{\texttt{\{}}\nsnt{graph\_descr}\nst{\texttt{\}}}
  }
  \nsdef{graph\_descr}{%
    \nsnt{id}\nst{;}\nsnt{graph\_descr} \nsorl
    \nsnt{id}\nst{-}\nst{>}\nsnt{id}\nsorl
    \nsnt{style\_descr}\nsnt{graph\_descr}\nsorl
    \nst{\epsilon}
  }
  \nsdef{style\_descr}{%
    \nst{[}\nst{style}\nst{=}\nsnt{style}\nst{]}\nst{;}
  }
  \nsdef{style}{%
    \nst{solid} \nsor
    \nst{dotted}
  }
  \nsdef{id}{%
    \nst{``}\nsnt{underscores}\nsnt{ident}\nst{``}
  }
  \nsdef{underscores}{%
    \nst{\_}\nsnt{underscores} \nsor
    \nst{\epsilon}
  }
\end{ngrammar}

The \textit{ident} identifiers are tokens containing English
characters and possibly integers. 
We parse this 
representation of the call-graph information using
XSB's \textit{DCG}s (Definite Clause Grammars).
We also implemented a tokenizer using the generic scanner in XSB-Prolog for
recognizing Modula-3 and Java programs\footnote{Available in
  \url{www.cs.stonybrook.edu/~shadjichrist/scanner.P}}. This scanner
splits the input in a list of token positions, and passes that
information to the parser. 
As we use the grammar to parse each edge represented in the 
\texttt{.eg} file, we use the representation encoded
to assert \texttt{edge/2} facts for each edge.
These facts have the format \texttt{
  edge(File1':'Source,File2':'Dest).} 

Within the same source file, the variables \texttt{File1} and
\texttt{File2} will be bound to the same atom, which is the name of
the source file. However, when all these edges are asserted, it is
useful to distinguish between different file names when we implement
the transitive closure. Our parser is a direct translation of
the above grammar, and can be found below:

\begin{lstlisting}
graph(File) --> ['digraph'], ['callgraph'], ['{'], graph_descr(File),
  ['}'].

graph_descr(File) --> identifier(_), [';'], graph_descr(File)
        | identifier(Source), ['-'], ['>'], identifier(Dest),
        { assert(edge(File':'Source,File':'Dest)) },
        style_descr, graph_descr(File)
        | [].

style_descr --> ['['], ['style'], ['='], style, [']'], [';'].

style --> ['solid']
        | ['dotted'].

identifier(Id) --> ['"'], underscores, [ident(Id)], ['"'].

underscores --> ['_'], underscores
        | [].
\end{lstlisting}

After each \texttt{.eg} file is processed, appropriate \texttt{edge/2}
facts will have been asserted into memory, giving us a representation
of the graph of the transitive closure, which Prolog can
easily process. The \texttt{parse\_files/0} predicate handles parsing the
\texttt{.eg} files; the \texttt{edge/2} assertion happens during
parsing, as shown in the above code (\texttt{scan\_file/2} is
implemented in the scanner library mentioned earlier).

Since we are asserting information gathered from the entire Linux
Kernel, one can imagine that the number of \texttt{edge/2} facts that
lies in memory is quite large, so it is useful to check how much time
it takes to assert all these data into memory. The predicate
\texttt{qtime/2} calculates the time a given query needs to
be computed, so the appropriate call below gives the time needed:

\begin{lstlisting}
|?- qtime(parse_files,T).
T = 104.2960
\end{lstlisting}

Perhaps unsurprisingly, it takes almost 2 minutes to assert all the
edges into memory. To see how many of these facts are being asserted,
a call to \texttt{findall/3} can be used:

\begin{lstlisting}
|?- findall(_,edge(_,_),L),length(L,N).
N = 52955;
\end{lstlisting}

We see that approximately 53,000 edges are being asserted into
memory. Although the assertion time may appear high at first glance,
XSB-Prolog offers the solution to such
problems by the means of advanced \textit{indexing} techniques, including
\textit{trie indexing}
\cite{sagonas1994xsb}. By adding an \texttt{:- index(edge/2,trie).}
directive, an index is created for the \texttt{edge/2} facts which are
now asserted much faster, as \texttt{qtime/2} divulges:

\begin{lstlisting}
|?- qtime(parse_files,T).
T = 19.6910
\end{lstlisting}

As a result of using indexing, we have reduced the time of needed to
assert the data-to-process in memory by a factor of more than 5. The
next step is to encode the transitive closure, and check its performance.

\begin{lstlisting}
:- table reachable_full/2.
reachable_full(S,D) :- edge(S,D).
reachable_full(File1':'Source,File2':'Dest) :-
        reachable_full(File1':'Source,_':'Temp),
        reachable_full(_':'Temp,File2':'Dest).  
\end{lstlisting}

With relevant calls to
\texttt{findall/3} and \texttt{qtime/2} we can find out the total
number of \texttt{reachable\_full/2} edges the transitive closure of
the graph contains, and how much time is needed to run through the
graph: 

\begin{lstlisting}
|?- findall(_,reachable_full(_,_),L),length(L,N).
N = 571295

|?- qtime(reachable_full(S1':'F1,S2':'F2),T).
T = 10.4210
\end{lstlisting}

The total size of the transitive closure of the initial graph is
roughly 570,000 edges, which is 10 times the number of the original
graph, and it takes about 10 seconds to go through the transitive
closure of the original graph. 

An interesting characteristic of this
implementation is that as users issue more queries,
future queries will be answered more quickly.
This is because additional queries will 
populate the \texttt{reachable\_full/2} table,
which effectively memoizes the results for future queries and can be
checked in constant time. 
For example, \texttt{kmalloc()} is a widely-used 
function within the Linux Kernel, so presenting some information
about it gives us a rough estimate of how long will it take for large enough
queries to be completed:

\begin{lstlisting}
|?- findall(_,reachable_full(_':'_,_':'kmalloc),L),length(L,N).
N = 8032

|?- qtime(T,reachable_full(_':'_,_':'kmalloc)).
T = 4.8000

|?- qtime(T,reachable_full(_':'_,_':'kmalloc)).
T = 0.0000  
\end{lstlisting}

The first query of all the functions that call
\texttt{kmalloc()} takes almost 5 seconds; yet the second query 
is effectively instant (constant time).
What is remarkable about this
behavior is that we get it in XSB-Prolog by just using the \texttt{:-
  table reachable\_full/2} directive.  Finally, had we not indexed the
\texttt{edge/2} facts, a call to go through the transitive closure of
the original graph would have been again a factor of 5 slower, as the
following query reveals:

\begin{lstlisting}
| ?- qtime(T,reachable_full(_,_)).
T = 65.1190
\end{lstlisting}

\section{An Interface for Users}
The last component of this framework is an interface between the
engine, described in the previous 2 sections, and users.
Rather than requiring users to install XSB-Prolog and issue
Prolog queries directly, 
we created a \texttt{PHP} website for users to issue queries to the database
and to display the answers. Our tool provides 4 different pre-compiled
queries to the transitive closure of the \texttt{edge/2} relation:
\begin{enumerate}
\item Provided a \textit{filename}, retrieve function call dependencies within
  the filename
\item Provided a \textit{source} function name, retrieve the names of all the
  functions that are called (directly and indirectly) from it
\item Provided a \textit{destination} function name, retrieve the
  names of all the functions that call it (directly and indirectly)
\item Provided a function name and a (possibly empty) list of other
  functions, retrieve the names of all functions that the former calls
  which are not contained in the later (this is also called a
  \textit{cut-off})
\end{enumerate}

\subsection{Pre-compiled Queries}
Since we decided that the results of these queries will be provided to
users via a web-browser, the output should be a specific HTML string
that represents the information in an understandable manner. The
\texttt{write\_html/4} predicate takes as arguments the name of the
operation (\texttt{file}, \texttt{source}, \texttt{dest} or
\texttt{cut\_off} as described in the list above), the
respective filename or source/destination/cut-off function names, and
in the cut-off case a (possibly empty) list of functions we have
already implemented. The fourth argument will be eventually bound to
specific HTML string that corresponds to the information we wish each
query to provide.

\subsection{The Server, Client and Web Interface}
At this point, we have all of the infrastructure needed to retrieve
function call-dependency information from the Linux Kernel, process it
and present the results in a user-friendly manner. The only pieces
needed are a server that will receive query requests from clients, a
\texttt{PHP} website that will be the interface between the users and the
engine, and a client program that will communicate the necessary
requests to the server, receive the answers and present them to the
user. Both the server and client are written in XSB-Prolog, and the
communication between them is implemented using the \texttt{socket.P}
library. 

This server code is included in the same source file as all the code
we presented above; it is necessary that the \texttt{edge/2} relation
and its transitive closure are kept in the server's memory
at-all-times if we want to take advantage of the tabling capabilities
of XSB. Calling the \texttt{server/0} predicate will initialize the
server and keep it running forever listening for requests from
clients. Once a request is received from a socket, the respective goal
is called, an answer list is constructed with an appropriate call to
\texttt{findall/3} and is returned to the client via another socket.

We call the \texttt{client/1} predicate with an
appropriate argument, which will be exactly the goal we want the
server to call and give us answers for. This goal is constructed
on-the-fly inside the \texttt{PHP} script located at
\url{http://ewl.cewit.stonybrook.edu/spyros/kernel.php}, based on the
users' selections. This script defines the various options users have
for querying the server. According to which option the user chooses,
and which arguments she provides, a query string is built on-the-fly, which
calls the client code mentioned above. In turn, the client code
communicates the request to the server, receives the answer when it
is computed and presents the result to the user.

\section{A Note on Efficiency}\label{eff}
This section presents running times and memory
consumption for specific queries to our engine. These statistics 
demonstrate the necessity of tabling and indexing in the backing Prolog engine,
even for medium-sized data analysis.  The \textbf{Query} column
lists the query in question; the \textbf{Tabling} and
\textbf{Indexing} columns list tabled and indexed
predicates, respectively, that we are interested in (and No if none is tabled);
the \textbf{Time} column lists the time needed to compute the query in seconds; and
the \textbf{Memory} column lists the space allocated at the end of the
computation in megabytes. 
We compute the memory consumption using XSB's
\texttt{statistics/0} predicate, by subtracting the memory consumption
when XSB is initialized (approximately 1MB) by the memory
consumption in the end of every computation.
A '-' entry in the table means that for the
particular query that option is not meaningful. The ``default'' entry
in the \textbf{Indexing} column means that trie indexing was not
enabled for the \texttt{edge/2} facts, hence first-argument indexing
was used, which is the default indexing mechanism XSB uses for all terms.

Table \ref{tab:t1} shows information regarding parsing and walking
the graph and its transitive closure. Tables \ref{tab:t2} and
\ref{tab:t3} show information regarding running the query 
\texttt{q1} (see Table \ref{tab:leg}). In Table \ref{tab:t2}, only a call
to \texttt{parse\_files/0} was made before taking 
the measurements, whereas in Table \ref{tab:t3}, calls to \texttt{infor} and
\texttt{infoe} queries (see Table \ref{tab:leg}) were made before taking the
measurements. 

\begin{table}[h]
\centering
\begin{tabular}{c c c r r }
  \textbf{Query} & \textbf{Tabling} & \textbf{Indexing} &
  \multicolumn{1}{c}{\textbf{Time}} & \multicolumn{1}{c}{\textbf{Memory}} \\
   &  &  &
  \multicolumn{1}{c}{\textbf{(s)}} & \multicolumn{1}{c}{\textbf{(Mb)}} \\
  \hline
  \texttt{parse\_files/0} & - & default & 111.8580 & 26.89 \\
  \texttt{parse\_files/0} & - & \texttt{edge/2},trie & 19.3700 & 22.32 \\
  \texttt{infoe} & - & default & 0.0170 & 22.32 \\
  \texttt{infoe} & - & \texttt{edge/2}, trie & 0.0170 & 22.32 \\
  \texttt{infor} & No & default & 71.6620  & 131.02 \\
  \texttt{infor} & No & \texttt{edge/2}, trie & 14.5410  & 127.11 \\
  \texttt{infor} & \texttt{reachable\_full/2} & default & 67.2490  & 131.17 \\
  \texttt{infor} & \texttt{reachable\_full/2} & \texttt{edge/2}, trie & 14.2860 & 127.07 \\
  \multirow{2}{*}{\texttt{infor}} & \multirow{2}{*}{\texttt{reachable\_full/2}}
  & \texttt{edge/2}, trie & 14.3530 & 127.71 \\ 
  & & \texttt{reachable\_full/2}, 1+2 & & \\
\end{tabular}
  \caption{Parsing and Walking through the graph}
  \label{tab:t1}
\end{table}

\begin{table}[h]
\centering
\begin{tabular}{c c c r r }
  \textbf{Query} & \textbf{Tabling} & \textbf{Indexing} &
  \multicolumn{1}{c}{\textbf{Time}} & \multicolumn{1}{c}{\textbf{Memory}} \\
   &  &  &
  \multicolumn{1}{c}{\textbf{(s)}} & \multicolumn{1}{c}{\textbf{(Mb)}} \\
  \hline
  \texttt{q1} & No & default & 131.0560 & 97.11 \\
  \texttt{q1} & \texttt{reachable\_full/2} & default & 121.0490 & 97.23 \\
  \texttt{q1} & No & \texttt{edge/2}, trie & 15.8320 & 92.97 \\
  \texttt{q1} & \texttt{reachable\_full/2} & \texttt{edge/2}, trie & 15.6110 & 92.55 \\
  \multirow{2}{*}{\texttt{q1}} & \multirow{2}{*}{\texttt{reachable\_full/2}}
  & \texttt{edge/2}, trie & 15.2670 & 92.52 \\ 
  & & \texttt{reachable\_full/2}, 1+2 & & \\
\end{tabular}
\caption{Query time/memory information without calls to \texttt{infor}
and \texttt{infon}}
\label{tab:t2}
\end{table}

\begin{table}[h!]
\centering
\begin{tabular}{c c c r r }
  \textbf{Query} & \textbf{Tabling} & \textbf{Indexing} &
  \multicolumn{1}{c}{\textbf{Time}} & \multicolumn{1}{c}{\textbf{Memory}} \\
   &  &  &
  \multicolumn{1}{c}{\textbf{(s)}} & \multicolumn{1}{c}{\textbf{(Mb)}} \\
  \hline
  \texttt{q1} & No & default & 57.0660 & 135.75 \\
  \texttt{q1} & \texttt{reachable\_full/2} & default & 64.1570 & 135.65 \\
  \texttt{q1} & No & \texttt{edge/2}, trie & 4.8150 & 131.84 \\
  \texttt{q1} & \texttt{reachable\_full/2} & \texttt{edge/2}, trie & 5.7140 & 131.25 \\
  \multirow{2}{*}{\texttt{q1}} & \multirow{2}{*}{\texttt{reachable\_full/2}}
  & \texttt{edge/2}, trie & 4.9440 & 131.80 \\ 
  & & \texttt{reachable\_full/2}, 1+2 & & \\
\end{tabular}
\caption{Query time/memory information after calls to \texttt{infor}
and \texttt{infon}}
\label{tab:t3}
\end{table}

\begin{table}[h!]
\centering
\begin{tabular}{l l}
  \texttt{infoe} & \texttt{findall(\_,edge(\_,\_),List), length(List,N)} \\
  \texttt{infor} & \texttt{findall(\_,reachable\_full(\_,\_),List), length(List,N)} \\
  \texttt{q1} & \texttt{findall(\_,write\_html(dest,kmalloc,\_,Ans),L)} \\
\end{tabular}
\caption{Legend}
\label{tab:leg}
\end{table}

The results presented in the tables above show some interesting facts
regarding the suitability of Prolog systems to handle and process
medium-sized datasets. Despite the fact that the Prolog code we wrote
for developing this tool is simple, compact and easy to maintain, had
XSB not offered tabling and indexing mechanisms, it would be
unsuitable for processing even medium-sized graphs. Indexing speeds-up
the time needed to parse and assert the facts by a factor of at least
5 in each case, and query processing time by a factor of almost
8. Tabling doesn't make any noteworthy changes to the speed of
answering a query for the first time, but it reduces the effort of
retrieving an answer already computed to a \textit{constant}
time. Moreover, if we pre-run some queries like \texttt{infoe} and
\texttt{infor} before launching the server, building queries for the
first time is even faster. 

Finally, memory consumption is not an issue for this tool.
Commodity systems often have several GB of RAM,
and  our tool used at most ~140MB when processing a
graph of more than half a million edges.

\section{Conclusion and Future Work}
In this paper, we investigated modern Prolog
systems for building tools able to handle data representing
medium-sized graphs. Our
first such tool is used to retrieve function dependencies from the Linux
Kernel to help developers understand complex attributes of the system.
The methods we described in the previous
sections are not only applicable to the Linux kernel; as a matter of
fact, \textit{any} C code base can be the subject of the function
dependency analysis we described. Moreover, we have
shown that tabling and indexing play an \textit{integral} role in such
efforts, thus making XSB-Prolog suitable for applications that were
generally out of Prolog's ``sweet spot''. All the code described in
this paper, along with the appropriate dataset obtained by compiling
the Linux Kernel can be found in
\url{http://www.cs.stonybrook.edu/~shadjichrist/kerfdep.tar.bz2}.

There are many directions for future work based on this tool. Among
the most challenging, would be figuring out the appropriate queries
that can help automating the process of modularizing
chunks of kernel code, instead of relying on a human expert. A quite
straightforward query we can write in our current setting, would be
one that includes in the \textit{cut-off} functions that are being
called from a particular one, but don't call other functions in their
bodies. In this query, we are making the assumption that functions
that do not have function calls in their bodies are implementing core
functionalities in the Linux Kernel, and are particularly written this
way for efficiency reasons. Although this can be a good starting point
for automating the process of modularizing code in the kernel,
more complex analyses should be used to get as close to what the human
expert would decide as possible.

One other
possible path would be the integration with popular source code
editors (such as Emacs, Vim, Eclipse), which would enable kernel
developers to use it more easily. With the
use of \textit{Tabling with Answer Subsumption} 
\cite{swift2010tabling} our framework is able to easily answer quantitative queries
regarding function dependencies, like for example ``which is the most
heavily used function in the Linux Kernel'', or ``which is the
function mostly called from function A''. Having this kind of
statistics in hand, will enable kernel developers to focus their
optimizations to particular heavily used functions and
components. Finally, larger datasets which contain much richer
information regarding the kernel's behavior (like for example LXR's
dataset) can be used. 

\section{Acknowledgments}

We thank the anonymous reviewers for their insightful comments
on earlier versions of this paper.  This research was supported in part by 
NSF CAREER grant CNS-1149229, NSF CNS-1161541, NSF CNS-1228839, 
and the Office of the 
Vice President for Research at Stony Brook University.

\bibliographystyle{splncs}
\bibliography{kerfdep}

\begin{thebibliography}{10}

\bibitem{colmerauer1996birth}
Colmerauer, A., Roussel, P.:
\newblock {The birth of Prolog}.
\newblock In: History of programming languages---II, ACM (1996)  331--367

\bibitem{covington1994natural}
Covington, M.A.:
\newblock {Natural Language Processing for Prolog programmers}.
\newblock Prentice Hall Englewood Cliffs (NJ) (1994)

\bibitem{rao1997xsb}
Rao, P., Sagonas, K., Swift, T., Warren, D.S., Freire, J.:
\newblock {XSB: A system for efficiently computing well-founded semantics}.
\newblock In: Logic Programming And Nonmonotonic Reasoning.
\newblock Springer (1997)  430--440

\bibitem{saha2006}
Saha, D.:
\newblock {Incremental evaluation of tabled logic programs}.
\newblock PhD thesis, Stony Brook, NY, USA (2006) AAI3258884.

\bibitem{saha2005symbolic}
Saha, D., Ramakrishnan, C.:
\newblock {Symbolic Support Graph: A space efficient data structure for
  incremental tabled evaluation}.
\newblock In: Logic Programming.
\newblock Springer (2005)  235--249

\bibitem{merritt1989building}
Merritt, D.:
\newblock {Building expert systems in Prolog}.
\newblock Springer-Verlag New York, USA (1989)

\bibitem{laera2004sweetprolog}
Laera, L., Tamma, V., Bench-Capon, T., Semeraro, G.:
\newblock {SweetProlog: A system to integrate ontologies and rules}.
\newblock In: Rules and Rule Markup Languages for the Semantic Web.
\newblock Springer (2004)  188--193

\bibitem{papadakis2011proton}
Papadakis, N., Stravoskoufos, K., Baratis, E., Petrakis, E.G., Plexousakis, D.:
\newblock {PROTON: A Prolog Reasoner for Temporal ONtologies in OWL}.
\newblock Expert Systems with Applications \textbf{38}(12) (2011)  14660--14667

\bibitem{stickel1988prolog}
Stickel, M.E.:
\newblock {A Prolog technology theorem prover: Implementation by an extended
  Prolog compiler}.
\newblock Journal of Automated reasoning \textbf{4}(4) (1988)  353--380

\bibitem{watson}
Lally, A., Prager, J.M., McCord, M.C., Boguraev, B., Patwardhan, S., Fan, J.,
  Fodor, P., Chu-Carroll, J.:
\newblock {Question analysis: How Watson reads a clue}.
\newblock IBM Journal of Research and Development \textbf{56}(3) (2012) ~2

\bibitem{huang2011datalog}
Huang, S.S., Green, T.J., Loo, B.T.:
\newblock Datalog and emerging applications: an interactive tutorial.
\newblock In: Proceedings of the 2011 ACM SIGMOD International Conference on
  Management of data, ACM (2011)  1213--1216

\bibitem{dlv}
{DLV}:
\newblock \url{http://www.dlvsystem.com/company}

\bibitem{semmle}
{SEMMLE}:
\newblock \url{http://www.semmle.com/}

\bibitem{ramakrishnan2007xcellog}
Ramakrishnan, C., Ramakrishnan, I., Warren, D.S.:
\newblock Xcellog: A deductive spreadsheet system.
\newblock Knowledge Engineering Review \textbf{22}(3) (2007)  269--280

\bibitem{becker2006secpal}
Becker, M.Y., Fournet, C., Gordon, A.D.:
\newblock Secpal: Design and semantics of a decentralized authorization
  language.
\newblock In: Proc. IEEE Computer Security Foundations Symposium. (2006)

\bibitem{sagonas1994xsb}
Sagonas, K., Swift, T., Warren, D.S.:
\newblock {XSB as an Efficient Deductive Database Engine}.
\newblock In: Proceedings of the ACM SIGMOD International Conference on the
  Management of Data, Citeseer (1994)  442--453

\bibitem{ext3}
Tweedie, S.:
\newblock Ext3, journaling filesystem.
\newblock
  \url{http://olstrans.sourceforge.net/release/OLS2000-ext3/OLS2000-ext3.html}

\bibitem{TxF}
Olson, J.:
\newblock {Enhance Your Apps With File System Transactions}.
\newblock {MSDN Magazine} (July 2007)
  http://msdn2.microsoft.com/en-us/magazine/cc163388.aspx.

\bibitem{spillane09fast}
Spillane, R., Gaikwad, S., Chinni, M., Zadok, E., Wright, C.P.:
\newblock {Enabling Transactional File Access via Lightweight Kernel
  Extensions}.
\newblock In: \fast{}. (2009)  29--42

\bibitem{googlefs}
Ghemawat, S., Gobioff, H., Leung, S.T.:
\newblock {The Google file system}.
\newblock SOSP (2003)

\bibitem{btrfs}
McPherson, A.:
\newblock A conversation with {C}hris {M}ason on btrfs: the next generation
  file system for {L}inux.
\newblock
  \url{http://www.linuxfoundation.org/news-media/blogs/browse/2009/06/conversation-chris-mason-btrfs-next-generation-file-system-linux}

\bibitem{mckenney04rcu}
McKenney, P.E.:
\newblock Exploiting Deferred Destruction: An Analysis of Read-Copy Update
  Techniques in Operating System Kernels.
\newblock PhD thesis, Oregon Health and Science University (2004)

\bibitem{swift2010tabling}
Swift, T., Warren, D.S.:
\newblock {Tabling with answer subsumption: implementation, applications and
  performance}.
\newblock In: Logics in Artificial Intelligence.
\newblock Springer (2010)  300--312

\end{thebibliography}

\end{document}